# Light Collimation and Focussing by a Thin Flat Metallic Slab

S. Ioanid, Ming Bai, N. García

*Laboratorio de Física de Sistemas Pequeños y Nanotecnología,*

*Consejo Superior de Investigaciones Científicas, Serrano 144 Madrid 28006, Spain*

A. Pons, P. Corredera

*Instituto de Física Aplicada,*

*Consejo Superior de Investigaciones Científicas, Serrano 144 Madrid 28006, Spain*

We present experimental and theoretical work showing that a flat metallic slab can collimate and focus light impinging on the slab from a punctual source. The effect is optimised when the radiation is around *the bulk, not at the surface, plasma frequency*. And the smaller the imaginary part of the permittivity is, the better the collimation. Experiments for Ag in the visible as well as calculations are presented. We also discuss the interesting case of the Aluminium whose imaginary part of the permittivity is very small at the plasma frequency in UV radiation. Generalization to other materials and radiations are also discussed.

Recently with the study of left-handed materials (LHM) [1], there has been a considerable works for focussing beams of electromagnetic radiation. Proposals of lenses that may increase resolution in the near field using LHM [2] have been made and the study of these possibilities has also been discussed [3]. On the other hand, photonic right handed materials exhibit all kind of effects, depending on what point of their band structure is explored by the impinging radiation [4-8]. Analogously because of the wave character of light, similar examples can be found with structured materials using acoustic waves [9-11]. Even focussing using plane lenses of finite size due to diffraction of multiple scattering by the lens edges has been discussed [11]. However, what have not been discussed to our knowledge are the collimating and focussing properties or infinitely large metallic thin films. We mean infinitely large film whose extension is much larger than the extension of the incident beam, in order to prevent diffraction effects due to the edges [11]. However in all the structured and photonic materials, there is also the problem of defining a refraction index that has to be valid in all directions.

In this paper we present an interesting effect of a plane metallic film, based on the negative real part of the permittivity at frequencies below the plasma frequency $\omega_p$ of the metals or other materials. Because of that, the radiation impinging at frequencies below $\omega_p$ will be tunnelling, that is to say, will behave the same way as particles tunnelling, and this process has the property of collimating and focussing the incident beam of light.

We have investigated further this point with light beams experimentally and theoretically and proved that optimum beam collimation and focussing takes place when the frequency of the radiation is around $\omega_p$ and not at the surface plasma frequency. Experiments as well as calculations are done for Ag that has a well-defined plasma frequency in the visible and calculations for Al in the UV, which is an ideal material due to his small imaginary part of the permittivity.

Consider a thin metallic film as described in Fig.1a and s-polarization radiation, impinging on the film. To treat the problem in a more intuitive way, we explain the physics in terms of scattering of wave-particles. It can be shown that the problem is equivalent to a quantum mechanics problem using the Schroedinger´s equation with a potential:
$$V = -k_0^2 (\varepsilon - 1) \quad (1)$$
Where, $\varepsilon$ and $k_0$ are the permittivity and the wave vector respectively. The permeability is unity. For a metal we have for frequencies below $\omega_p$, $\varepsilon<0$. Therefore for radiation frequencies smaller than the plasma frequency, the problem is equivalent to a quantum mechanical tunnelling by a square well problem. Then it is clearly seen in Fig.1b, for the rays going perpendicular to the film the tunnelling probability is larger than for those impinging obliquely because those ones have to transverse longer distances in the metal, with $L_1>L_0$. This can be easily found for formulas in quantum mechanical textbooks [12]. According to that, when a radiation coming from a punctual source impinges the film, the normal rays will have more transmittivity than those oblique. As a consequence, the outgoing transmitted beam will be collimated, as illustrated in Fig. 1b. This can be physically understood as follows.

Consider Snell´s law for an incident wave impinging the surface, from the air, with angle $\theta$, then:
$$\sin \theta_m = \frac{\sin \theta}{\sqrt{\varepsilon}} \quad (2)$$

where $\theta_m$ and $\varepsilon$ are the angles of propagation and the permittivity of the metal, it is clear that for $\varepsilon$ going to zero, less and less angles of incidence are permitted and the trajectories in the metal will have longer angles and then distances for the light to cross the metal are much longer. As these trajectories are in tunnel, evanescing, the better focussing takes place where $\varepsilon$ tends to vanish. This is when there is only a real part in the permittivity. If there is a small imaginary part (case of Ag), then a little more plying has to be done but basically one arrives at the same conclusion. The optimal frequency for obtaining the effect is shifted slightly to lower energies depending of the magnitude of the imaginary part. It is a question of reasoning a little more and to study the trajectories when the light come out again in the air to see that there is a concentration of trajectories, higher intensity, near the surface in front of the point source. That can be interpreted as a focussing. The case of Al is perfect because its imaginary part is practically zero.

The above ideas are proved experimentally and with calculations for the Ag permittivity [13], as well as for the interesting case for the Al in the UV given the small imaginary part of the permittivity near plasma frequency [14]. The experiments were performed for a film of Ag deposited on a 0.15mm glass subtracts. A wide opened incident punctual source was obtained by focussing a laser beam with a cylinder lens, linearly s-polarized. Using a photon multiple tube (PMT) as detector, we measured the transmitted intensity in angular distribution, as illustrated in Fig.1b. The angular transmission distribution curves were later calibrated against the inhomogeneity of the incident laser beams.

One of the reasons we choose Ag as film component for experiments is that its plasma frequency falls near the visible range (325nm), and its dielectric constant has a small imaginary part (the plasma is well defined). We measured the collimating effect with lasers of 632.8nm, 488nm and 325nm in wavelength. The collimation is shown by the angular transmission distribution versus the outgoing angle in Fig 2a. The closer to the plasma frequency, the narrower is the intensity profile, which means the beam is more collimated.

We also performed transfer matrix calculations for the multi-layer system. In Fig. 2b, the calculations verified the experimental results that for the same film thickness, the shorter source wavelength is, the smaller the angular transmission distribution width is, as shown in Fig 2b, reaching a minimum around 325nm, the plasma frequency of the silver. But the width then grows for still shorter wavelengths, where according to formula (1) the light is no longer tunnelling in the film but propagating. The collimation is effectively optimised at the plasma frequency, and given the structure of the permittivity, the transmission is the largest (approximately 3%), presented in the inset of fig. 2b. The same regular pattern of having a maximum collimation around plasma frequency is also obtained with other thicknesses, such as 100nm and 250nm, not shown but confirmed by experiments and calculations.

In order to see focussing effects we have performed a simulation using Finite-difference Time-domain (FDTD) method that gives the field intensities near the surface. The FDTD method is constructed with Drude dispersion relation well fitted to Ag optical dispersion constants [13] in the frequency range we studied. As shown in fig. 2c for the plasma frequency, the 150nm thick Ag film confines the angular width of the transmission beam from 180 degree to less than 60 degree. Meanwhile, it can be seen a concentration of the intensity near the surface and in front of the source that indicates a focussing. More results for different energies and polarizations will be presented elsewhere.

With calculations, we noticed that the collimation effect is significantly improved by setting the imaginary part of the permittivity to be zero at plasma frequency. This interesting result draws our attentions on Aluminium who has an even smaller permittivity than Ag at plasma frequency (UV). In this case the collimating effect is more obvious, and the transmittivity is also higher than Ag case, as shown in Fig.3a. It will be very interesting to perform the experiments for Al. However, this present a non-negligible complication in experiment. To perform this experiment one should do it with an Al membrane hanging without substrate.

The ideas of collimation and focussing developed here can be extended to other materials with plasma frequencies at different regions: infrared, microwaves, etc. or oxidizing the metals, where the losses may be smaller and the plasma frequencies changed. It seems that its application can have a wide range of frequency just by adequately choosing the material and the incident radiation.

This work has been supported by the EU-FP6, Molecular Imaging Project and by the Spanish DGICyT.

**Figure Captions**

**Figure 1**

(a) Scattering geometry with energy dependent optical potential $V = -k_0^2(\varepsilon - 1)$, with incident frequency $\omega = -ck_0^2$.

(b) The experimental scheme for measuring the transmitted intensity from a metal film in angular distribution, with a PMT detector. The Ag film was deposited on a glass subtracts. An s-polarized laser beam was focused on one side of the sample slab by a cylinder lens to produce a punctual source.

**Figure 2**

Angular transmission distribution versus the outgoing angle through a 150nm Ag deposited on a 0.15mm thick glass substrate.

(a) Experimental transmission with different incident lasers 633nm, 488nm and 325nm in wavelength. The horizontal line shows the angular distribution for the incident intensity of the point source.

(b) Calculated transmission results with different incident lasers going from 633nm to 280nm wavelengths. For comparison, the result of a 250nm thick Ag film incident at 325nm (plasma frequency) (open circles) laser is also presented. The inset image represents the corresponding transmission intensity and the Ag optical constant versus wavelengths for the Ag film.

(c) FDTD simulation of the collimation effect of a 150nm Ag film at its plasma frequency (3.8eV, 325nm). Intensity distribution was log scaled. Notice the larger intensity near the surface in front of the source.

**Figure 3**

Calculated angular transmission distribution versus the outgoing angle through a 150nm Aluminium thin film, with different incident lasers wavelength ranging from 633nm to 75nm. For 75nm, above the plasma frequency (83nm), where the angular spread is wider than 83nm, because the wave does not tunnel but propagates in the film. The inset image represents the corresponding transmission intensity and the Aluminium optical constant versus wavelengths.

**Figures**

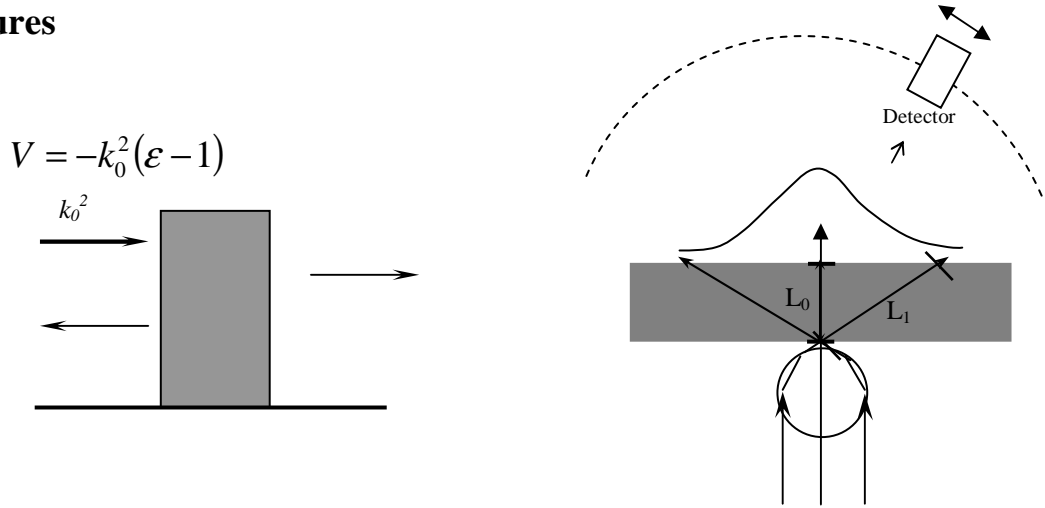

Figure 1. (a)  (b)

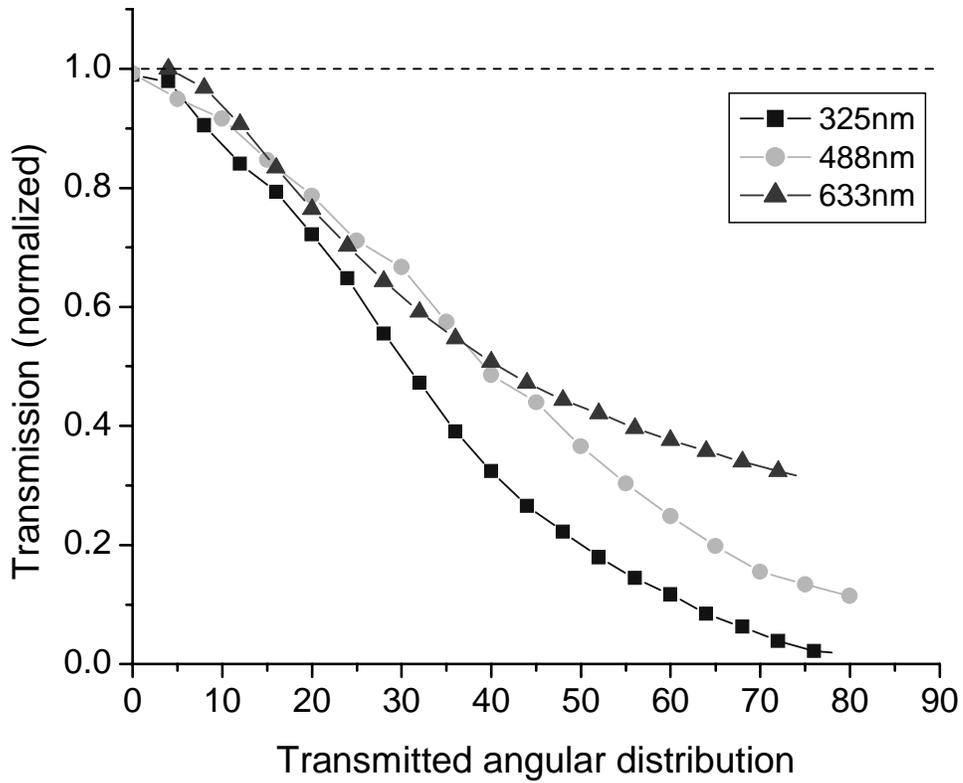

Figure 2. (a)

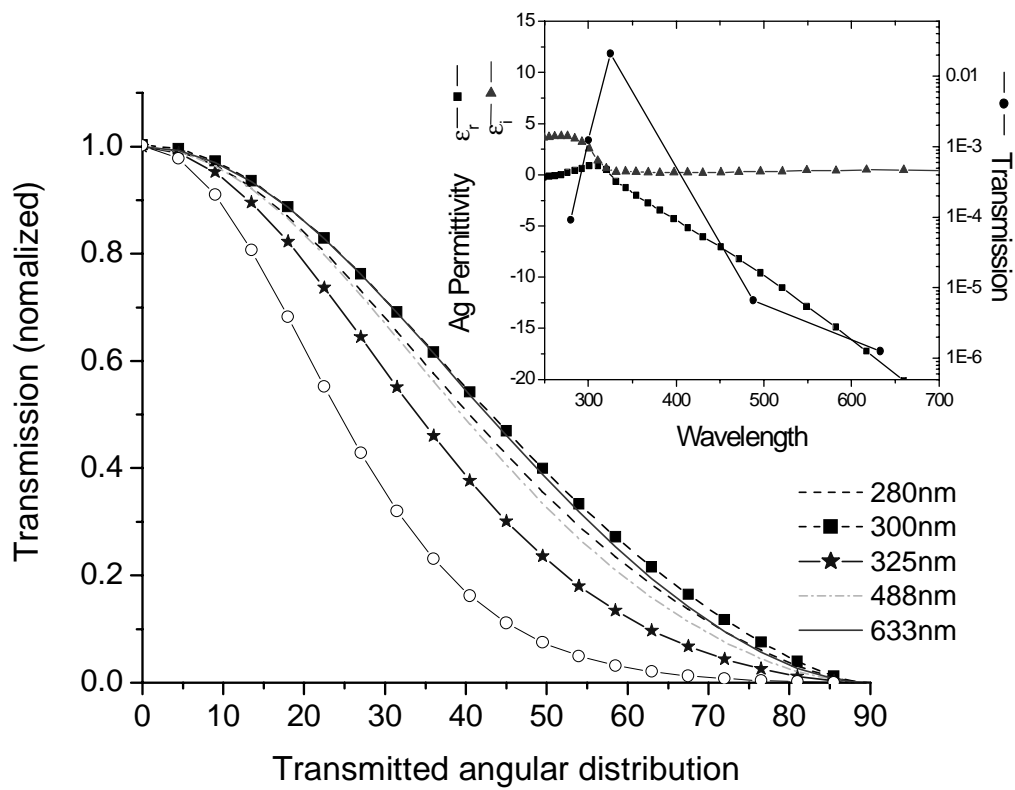

Figure 2 (b)

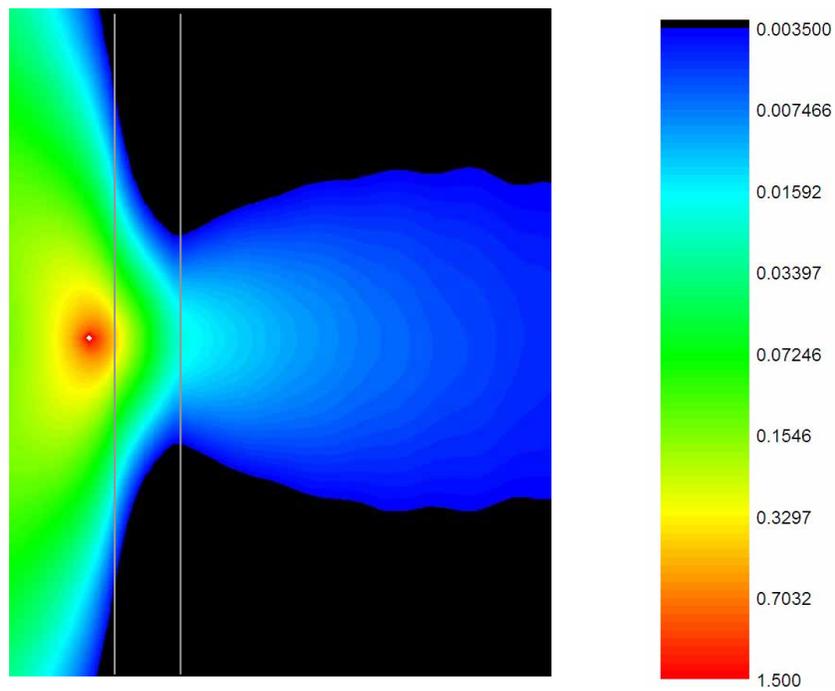

Fig.2 (c)

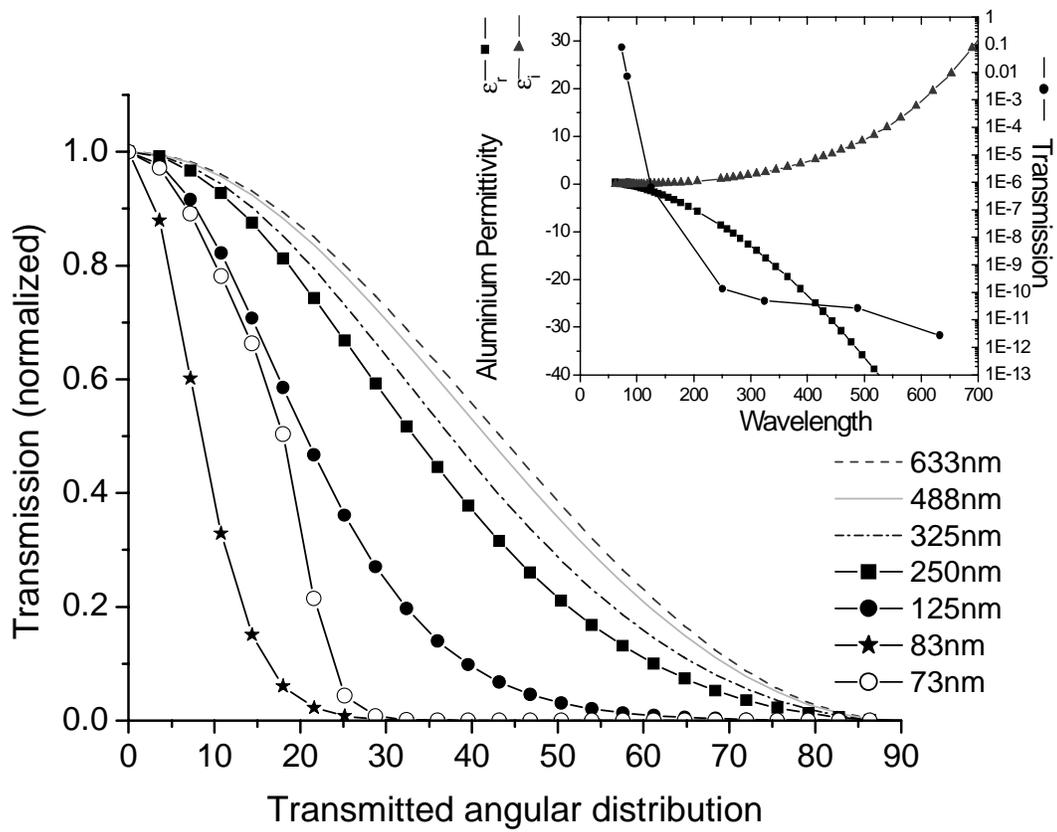

Figure 3.